# A NEW APPROACH IN PACKET SCHEDULING IN THE VANET

Sayadi Mohammad Javad[1] and Fathy Mahmood[2]

[1]E-Learning center, Iran University of Science and Technology
mjsayadi@vu.iust.ac.ir

[2]Department of computer engineering, Iran University of Science and Technology
mahfathy@iust.ac.ir

## ABSTRACT

*Vehicular Ad hoc Networks (VANET) are expected to have great potential to improve both traffic safety and comfort in the future. When many vehicles want to access data through roadside unit, data scheduling become an important issue. In this paper, we identify some challenges in roadside based data access. To address these challenges we first review some existing scheduling schemes. We then propose a priority scheduling and finally show that using this idea can increase QOS compare to previous algorithms.*

## 1.INTRODUCTION

Vehicular network applications require wireless mobile communications. Currently, there are several possible paradigms for wireless mobile communication, for example, cellular, ad hoc, wireless LAN, and Info-stations [5][6]. Clearly, the choice of technology depends on the application that the network is intended to support. For this reason we need to have a clear insight into these applications and their requirements.

Integrating a network interface, GPS receiver, different sensors and on-board computer gives an opportunity to build a powerful car-safety system, capable of gathering, processing and distributing information. Numerous applications can be deployed in a network established with such equipped vehicles and proper infrastructure.

These applications are either safety related or comfort related [3].

Safety related applications usually demand direct communication due to their delay-critical nature. One such application would be emergency notifications, e.g. emergency braking alarms. [2]

The general aim of comfort related applications is to improve passenger comfort and traffic efficiency.

That could include nearest POI (Points of Interest) localization, current traffic or weather information and interactive communication.[1]

Data dissemination model in a safety related application is a push model [4]. This model is based on the idea that nodes sensing of interest continuously "push" the information into the network. For example, a vehicle detecting a traffic jam situation would disseminate this information via the VANET.

Main communication model in a comfort related application is a pull model that is the traditional communication model in the Internet.[4] An application sends a request, the request is forwarded to a node at the destination and the destination node sends a reply including the requested information.

In some comfort related applications such as traffic warning or whether condition the communication model is a push. Comfort related applications must use roadside unit to transmit their data but safety related application information exchanged without requiring any fixed infrastructure.[7]

In section 2 we review some existing comfort related scheduling schemes. Afterward in section 3 priority scheduling algorithm is considered and finally in section 4, 5 we show that using this idea increase QOS compare to previous algorithms.

## 2. RELATED WORKS

Recently roadside-based on data communication has received considerable attention. Roadside units allow vehicles to access data stored on the roadside unit or access the Internet.[8]

Some comfort related application messages such as traffic messages or whether condition are broadcast by roadside unit to vehicles.[9](Figure1) In this communication model a roadside unit acts as a buffer and vehicles can upload/download their data on/from it.[10]

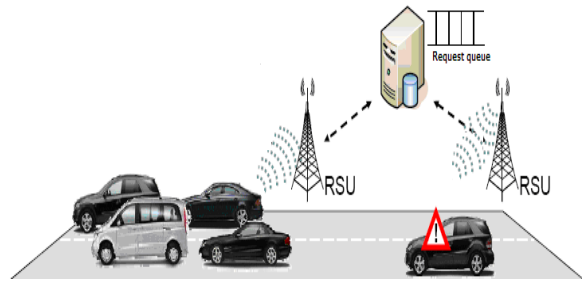

Figure1: Roadside based communication model

When vehicles enter the roadside area, send requests to the roadside unit if they want to access the data. All requests are queued and based on scheduling algorithm; the server serves one of them and removes it from queue.

Note that the requests are only active for a short period of time because vehicles are moving and they stay in the roadside unit area for a short period of time and requests not served within a time limit is dropped as the vehicles move out of the roadside unit area.

Now we survey solutions for scheduling roadside based data access.

Several scheduling algorithm proposed to roadside based communication model in VANET. In most of them, the metrics are "service ratio" and "data quality".

A good scheduling algorithm should serve as many requests as possible and update data in time and try to avoid data staleness.

"Data size" and "Deadline" are two parameters that used for these algorithms. Therefore each request is characterized by <V-id, D-id, Op-Type, Deadline>.[10]

In this 4-tuple, V-id is the identifier of the vehicle, D-id is the identifier of the requested data, Op-Type is the operation that the vehicle wants to do (download or upload) and Deadline is the critical time constraint of the request.

FDF[1] and SDF[2] are two naive schemes that in FDF, the request with the earlier deadline will be served first and in SDF, the request ask for a small size data will be served first.

As shown in Figure2 [10], if the request arrival rate is low, FDF acts better than SDF. Because when the workload is low, the deadline factor has more impact on the performance.

When the request arrival rate increases, SDF performs relatively better.

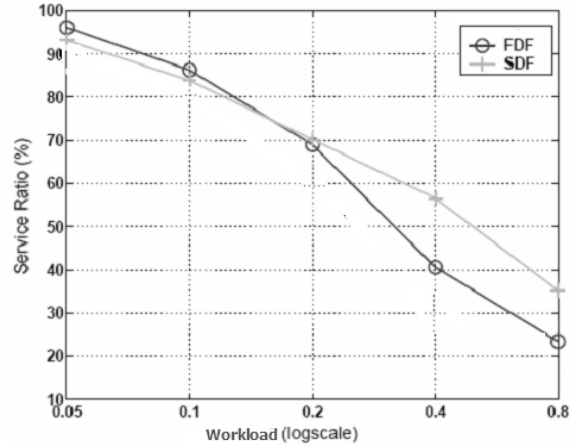

Figure2: Service ratio for FDF and SDF

A newer scheduling algorithm is D*S that integrate the deadline and data size to improve the performance of scheduling. In this scheme each request is given a service value based on its deadline and data size as its service priority weight. [11]

$$DS\_Value = (DeadLine - CurretntClock) \times DataSize$$

This algorithm computes DS_Value for all received requests and always serves the requests with the minimum DS_Value.

D*S scheduling algorithm considers both request deadline and data size but it serves one request at one time.

This algorithm improved by download optimization. If some vehicles request the same data to download, several requests may be served through a single broadcast. In this approach the data with the more pending request should be served first. [12][13]

This new scheme called D*S/N and the service value DSN_Value for that, is calculated as

$$DSN\_Value = (Deadline - CurrentClock) \times DataSize / Number$$

---

[1] First Deadline First
[2] Smallest Data size First

Note that in this equation deadline may be the earliest, the median or the mean deadline of the pending request group. Simulation results for these selecting deadlines are shown in Figure3 [10].
To improve the "Data quality" metric and maintain a balance between serving download and upload a Two-Step scheduling scheme was proposed.

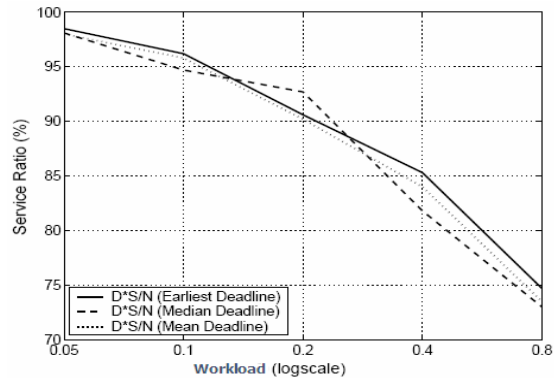

Figure3: Service ratio D*S/N algorithm

Simulation results show that the Two-Step scheduling scheme outperforms other scheduling schemes and is adaptive to different workload scenarios.

## 3. PACKET SCHEDULING USING MULTI LEVEL QUEUE

The roadside based communication model is developed to exchange the packets of comfort related and non real time application such as Internet access and instance messaging. All scheduling algorithms proposed in the roadside based structured suppose that all request packets have same type and there is no difference regarding their type.
D*S algorithm, for example, which is a basic algorithm in the roadside based communication model, uses a priority weight for selection of a request to get service. Priority weight is influenced only by deadline and data size parameters and that "which application produced request packet" and "required service level" are not important. In this section we show that previous scheduling algorithms will be improved to give service depend on type of application that produce the request packet.
In first step of algorithm improvement, we try to set deadline parameter with a small value, manually, regarding direct impact of request deadline parameter. This will give a higher chance to request packet to get service.
Although this improvement helps us to achieve to a better service to real time application request packet, some events could prevent us to achieve the goal.
A non real time application packet, for example, which does not need high service priority, can get a higher priority after a long waiting time comparing with real time application request packet priority. It will result in depriving real time application request of getting service.
It is because of the fact that deadline of packet cannot be smaller than a certain level. If the value of packet deadline be so small, it will be expired soon and removed from service queue. Next step in scheduling algorithm improvement is to add a new parameter which indicates the type of packet. The idea is to classify the request packets by a field called "TYPE". All request packets from application with same degree of importance, will get the same value for the field of "TYPE". Therefore, it will customize and differentiate services regarding made classes.
In roadside units several queues are considered. When a request packet arrives, it is put in a queue depends on its TYPE field value. (Figure4)

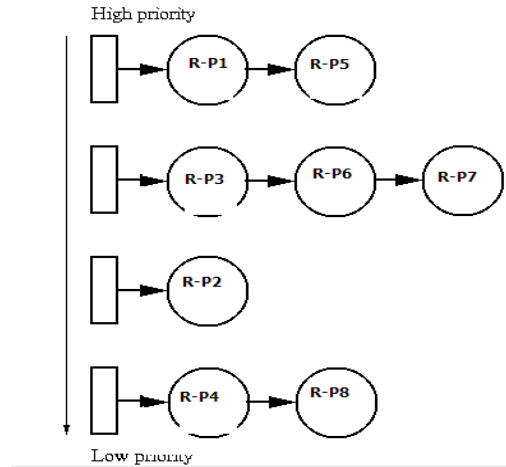

Figure4: Request packet queues in RSU

For example, request packet of real time applications are put in one queue and request packet of non real time applications are put in another queue.

In this scenario requests which are in higher priority queue, get service first and while there is a request packet in this queue, other request packets from other queue are not selected to get service. Selection of request packets inside the same queue can be done according to previous algorithm policies. In this improvement, by applying this classification some of packets never get chance to receive service, and their deadline will be expired before receiving service.

## 4. IMPLEMENTATION

In this section we evaluate the performance of the scheme proposed in section 3 in large scale network by means of GloMoSim simulator. In this simulation we fix the length of the road to 1000 meters with 2 lanes on each side. We use 802.11b as the wireless media with a transmition range of 250 meters.

During a simulation run, vehicles periodically send their request packets to the RSUs and request packets are classified in two priority classes. Therefore two priority queues are considered in RSUs. One of them is used to differentiate request packets with high degree of priority and another queue is used to maintain request packets with low degree of priority.

We use two link lists to implement priority queues.

| Communication | |
|---|---|
| Mobility Model | Tow-way ground |
| Mac Model | IEEE 802.11b |
| PHY Data Rate | 1Mbit/sec |
| Transmit Range | 1Km |
| Vehicular Traffic Model | |
| Road Length | 1km |
| Number of lanes | 2 per direction |
| Desired velocity | 140km/h |

Table1: simulation setup

## 5. SIMULATION RESULTS

In this evaluation we study the service ratio and Figure5, 6 show the simulation results. As we show in figure5, this algorithm works weaker than previous schemes in service ratio for non real time application request packets (low level request packets) most of removed request packets are from this class, because they receive service last. But figure6 shows that real time application request packets (high level request packets) receive a good service and this class of packets have a good service ratio.

Therefore, this scheme cannot provide acceptable service ratio for low level request class but the request packets in high level class received service very good.

In other words, it is natural that some of packets are removed because of their expiration of their deadline, but we can manage the probability of success of packets in receiving service or removing from queue.

If request packets which are arrived at RSU to receive service are classified, our scheme works better than previous scheduling algorithm. Because in other algorithms all of packet types have same chance to get service.

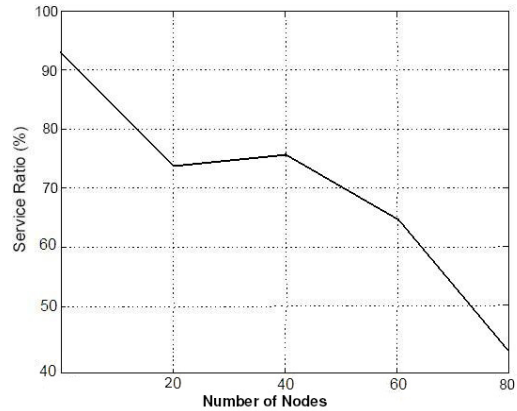

Figure5: service ratio for non real time application request packet

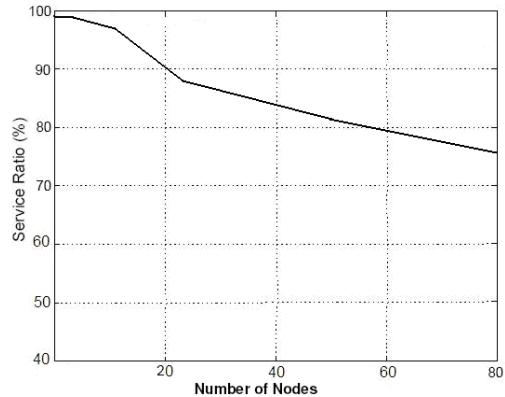

Figure5: service ratio for real time application request packet

Therefore by this approach we try to reduce probability of deleting request packet in important class and increase the network QOS.

## 6.CONCLUSION

In the paper we address several scheduling schemes in roadside based communication model in VANET. We also identify the effect of classification of request packets depend on their importance degree on service ratio and proposed a multi level queue scheduling algorithm. Simulation results show that this idea works better in important request packet class and help us to increase the QOS in the network.